\documentstyle[12pt,epsf]{article} 
\def\a{\alpha}

\def\b{\beta}
\def\w{\omega}

\input epsf
\begin{document}
\title{
 Topological Quantum Field Theories -- \\
A Meeting Ground for Physicists and Mathematicians
} 
 
\author{Romesh K. Kaul  \\
The Institute of Mathematical Sciences, \\
 Taramani, Chennai 600 113, India. \\
{\it Email:  kaul@imsc.ernet.in}}
\maketitle
 
\begin{abstract}
Topological quantum field theories can be used as  a powerful
tool to probe geometry and topology in low dimensions.
Chern-Simons theories, which are examples of such 
field theories, provide a field theoretic
framework for the study of knots and links in three
dimensions.  These  are rare examples of
quantum field theories which can be exactly 
(non-perturbatively) and
explicitly solved. 
Abelian Chern-Simons theory provides a field theoretic
interpretation of the linking and self-linking numbers
of a link. In non-Abelian theories, vacuum
expectation values of Wilson link
operators yield a class of polynomial link invariants; the
simplest
of them is the famous Jones polynomial. Other
invariants obtained  are more powerful than that of Jones.
Powerful methods for completely analytical and
non-perturbative computation of these knot
and link invariants have been developed.
In the process answers to some of the open problems
in  knot theory are  obtained.
From these  invariants  for unoriented and framed links 
in $S^3$, an invariant for any three-manifold can  
be easily constructed by exploiting the Lickorish-Wallace
surgery
presentation of three-manifolds. This invariant 
up to a normalization is
the partition function of the Chern-Simons field theory. 
Even perturbative analysis of the Chern-Simons theories are 
rich in their mathematical structure; these provide
a field theoretic interpretation  of Vassiliev knot invariants.
Not only in mathematics, Chern-Simons theories
find important applications in three and four dimensional
quantum gravity also.

\end{abstract}

\vskip1.5cm

\newpage
 
\section{Introduction}

Many a time advances in mathematics and physics have
occurred hand in hand. Newton's theory of mechanics and
the development of  techniques of calculus are a
classical example of this phenomenon. Another example is 
the developments in differential geometry inspired 
by  Maxwell theory of electromagnetism  and Einstein
theory of general relativity. A recent glorious
example is the developments of topological quantum field
theories and their relevance to the study of geometry
and topology of low dimensional manifolds.

The application of topological quantum field
theories reflects the enormous interest
generated both by mathematicians and field
theoreticians in building a link between quantum
physics through its path integral formulation on
one hand and geometry and topology of low
dimensional manifolds on the other. These are indeed
deep links which are only now getting explored. It does
appear that the properties of low dimensional
manifolds can be very successfully unraveled by
relating them to infinite dimensional manifolds
of fields. This provides a powerful tool to study
these manifolds notwithstanding the `lack of
mathematical rigour' in defining the functional
integrals of quantum field theory.
Indeed, an axiomatic formulation of topological
quantum field theories has also been attempted.

Toplogical quantum field theories are independent of the
metric of curved manifold on which these are
defined; the expectation value of the energy-momentum
tensor is zero, $\langle T_{\mu \nu} \rangle = 0$. 
These possess no local propagating degrees of freedom;
only degrees of freedom are topological. Operators  of interest
in  such a theory are also metric independent.

To illustrate how ideas of quantum field theory
can be used to study topology,
we shall  focus our attention here on 
recent important developments in  Chern-Simons
gauge field theory as a topological quantum field
theory on a three-manifold. This theory provides a field
theoretic framework for the study of knots and links
in a given three manifold\cite{ati} - \cite{kau5}.
It was A.S. Schwarz  who first conjectured \cite{sch} that
the now famous Jones polynomial \cite{jon} may be
related to Chern-Simons theory. E. Witten in his pioneering
paper about ten years ago demonstrated this connection\cite{wit}. 
In addition, he set up a general field theoretic framework to study
knots and links. Since
then enormous effort has gone into developing an exact
and explicit non-perturbative  solution of this field theory.
Many of the standard techniques of field theory find
applications in these developments. The interplay
between quantum field theory and knot theory has paid
rich dividends in both directions. Many of the open
problems in knot theory have found answers in the
process.

Wilson loop operators are the topological operators of
the Chern-Simons gauge field theory. Their vacuum 
expectation values are the topological invariants for 
knots and links which do not
depend on the exact shape, location or form of the
knots and links but reflect only their topological
properties. The power of this framework 
is so deep that it allows us to study these
invariants not only on simple manifold such as
three-sphere  but also on any arbitrary three-manifold. 

The knot and link invariants obtained from these
field theories are also
intimately related to the integrable vertex models in
two dimensions\cite{wad, kau5}. 
These invariants have   also been approached  in different
 mathematical frameworks. A quantum
group approach to these polynomial invariants has been
developed\cite{kiril}.  
Last decade or so has seen enormous activity in these
directions in algebraic topology.

A mathematically important development is that these
link invariants provide a method of obtaining
 a specific topological
invariant for three-manifolds\cite{wit,res}
in terms of invariants for framed unoriented
links in $S^3$\cite{lic, kau5, naik}.  In the following, 
we shall review these developments.

Not only in mathematics, Chern-Simons theory has also played a major
role in quantum gravity. Three-dimensional gravity with a negative
cosmological constant, itself a topological field theory,
can be described by two copies of $SU(2)$ Chern-Simons theory.
Even in four dimensional gravity, Chern-Simons
theories find application. For example, the boundary degrees 
of freedom of a black hole in four dimensions,
are described by an $SU(2)$  Chern-Simons field theory.  
This has  allowed an exact  calculation of  quantum entropy
of a non-rotating black hole.
The formula  so obtained for a Schwarzschild black hole, while 
agreeing with the  Bekenstein-Hawking formula
for large areas, goes beyond the semi-classical result.

Before explaining how a field theoretic
framework for knots and links can be developed, 
let us start with a brief discussion of knots and links.

\section{Knots and links: an elementary introduction}

{\it What is knot?} A smooth non-intersecting
closed curve in a three-manifold is a knot.
Oriented closed curves are oriented knots.
A string with its ends joined in the shape of a circle
without any
entanglements is a model for the simplest
non-intersecting closed curve called {\it
unknot}. With a given knot, 
we associate a {\it knot diagram} obtained by projecting the
knot on to a plane with a minimum number of
double points. In such a diagram over-crossings
and under-crossing are to be clearly marked.
The number of double points in a knot diagram
is called its {\it crossing number}.
A  few simple knots with low crossing numbers are:

\vskip0.5cm
\centerline{\epsfbox{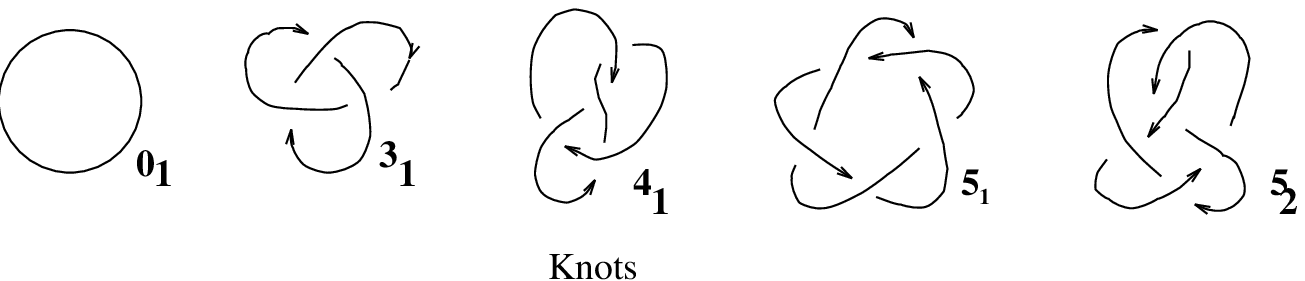}}
\vskip0.5cm

Clearly, for
a given minimum number of crossings, there can
be more than one type of  topologically inequivalent
knots. The number of knots increases rapidly
with the crossing number. For crossing number 9,
there are 49 knots (not distinguishing mirror
reflections), for 10 there are 165 and for
crossing number 11 we have 552 knots.
For 13 crossings, there are more than 10,000
different knots.

{\it What is a link?} A collection of a number
of
oriented non-intersecting loops (knots) is an
oriented link. A knot
then is single component link. 
Links like knots can be represented by their
two dimensional projections, {\it the link
diagrams} with minimum number of double points,
but
with the over-crossings and under-crossings
clearly marked. Examples of a few two-component links
are:
 
\vskip0.5cm
\centerline{\epsfbox{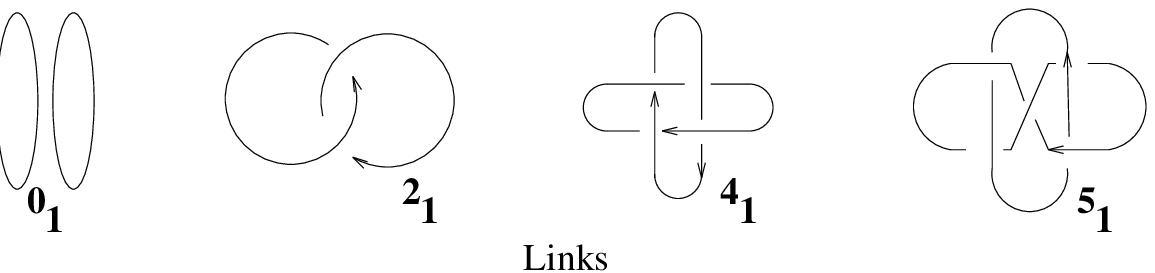}}
\vskip0.5cm

To a topologist, length, thickness or precise
shape of a  knot are not of any interest.
Two knots or links are to be identified if one can be
made to go continuously into other by shrinking
or stretching or wiggling without snapping the
string. There is a minimal set of elementary rules which
encode these qualitative notions more precisely. These
are the three  Reidemeister moves   which do not
change the  topological  type of a link:

\vskip0.5cm
\centerline{\epsfbox{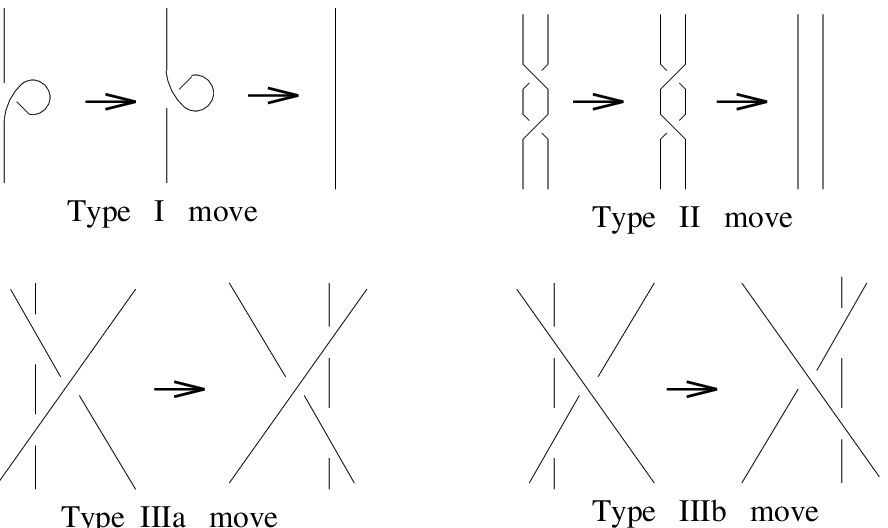}}
\vskip0.5cm
 
\noindent Invariance  under  all these three moves 
is called invariance under {\it ambient isotopy}. If a quantity
is invariant under type II and III moves only, but not under 
type I moves, it is said to be a {\it regular isotopic invariant.} 

The Reidemeister move III is of particular interest. It 
represents a defining relation for the generators
of braids. In addition, it is a graphical representation of the 
Yang-Baxter
relation of statistical mechanical models. These facts are  not
accidental but reflect
a deep connection that knots and links have with braids and
exactly solvable two-dimensional vertex models \cite{wad}. In fact
this connection has been successfully exploited to obtain
infinitely many new exactly solvable  statistical mechanical
models\cite{kau5}.

Though the Reidemeister rules 
are so simple, it is not an easy
exercise  in general to tell whether given two knots or
links are topologically distinct or not.  For example,
it took nearly eighty years, since the time of  knot
tables of C.N. Little from  the end of last century to the work of
K.A. Perko in 1972, to  recognize that
the knots in the figure below are isotopically
equivalent\cite{perko}:
 
\vskip0.5cm
\centerline{\epsfbox{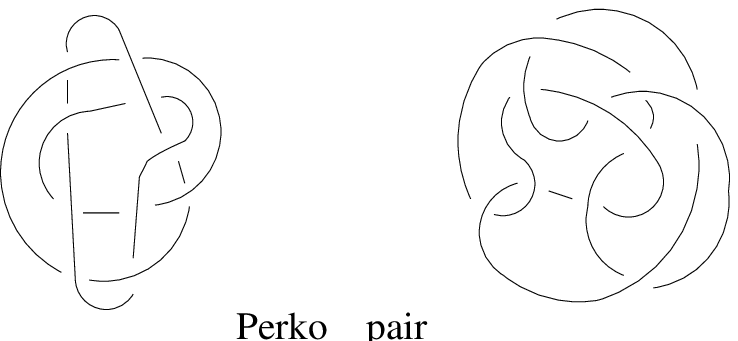}}
\vskip0.5cm

Finding mathematical methods for distinguishing
knots and links is indeed an important problem
in knot theory. To this end, some definite
invariants, called {\it link invariants}, are 
associated with the links. These are mathematical
expressions which depend only on the isotopic
type of the link and not on any of its particular
representations. Some such invariants are  in the
form of polynomials. First  polynomial invariant 
was  discovered in late twenties by J.W. Alexander\cite{alex}. 
It took almost sixty more years before the next
one was discovered by V.R.F. Jones\cite{jon}. The new invariant
proved to be topologically more powerful than
that of Alexander. For example, unlike Alexander
polynomial, Jones polynomial does distinguish 
many mirror reflected knots. Soon after, a two
variable generalization of Jones invariant was
found \cite{homfly}. Though two distinct Jones polynomials
do represent two isotopically distinct knots,
the converse is not always true. There are
examples of distinct knots with same Jones
polynomial. Still Jones' work represents a leap
forward in the developments of knot theory. What
is impressive about the topological field
theoretic description of knots is that it
provides a whole variety of link invariants in a
straight forward manner. Of these  Jones 
one-variable polynomial and its two-variable 
generalization are the simplest examples.

Before starting a discussion of knots and links
in terms of a quantum field theory, let us make a
few historical remarks about knots and links
in physics.

\vspace{0.5cm}

{\bf A few historical remarks:}~ Knots and links
first captured the imagination of physicists when
Lord Kelvin (William Thomas) introduced them as early as 1857
as fluid-mechanical models of atoms \cite{kel}. Reluctant
to accept the prevailing notion of an infinitely
rigid point-like atom, he thought of atoms as
vortex-lines in a perfect homogeneous fluid, the
{\it ether}. Different sorts of atoms were then
to differ in accordance with the number of
intersections of these vortex rings. ``Stability'' of the atoms
in this theory  thus is a reflection of the fact that knots do preserve
their essential knottedness during their movement.
Indeed Lord
Kelvin would have wanted to develop a new theory
of gasses, theory of elastic solids and liquids
based on the dynamics of these vortex atoms --
a programme he did not complete nor was
considered by later day physicists worth while
in this context. However, a new area, {\it knot
theory}, of mathematics was born. 

Two contemporary Scottish physicists, J.C.
Maxwell and P.G. Tait did find Lord Kelvin's
hypothesis attractive enough. Tait had hoped to
explain the position of  lines in the
spectrum of a chemical element  from the knot
type representing it. Thus, it was natural for
him to attempt the formidable task of
classifying knots in three-space. For this he
needed some measure of complexity of a knot.
Thus the concept of the {\it degree of
knottedness} was introduced. This is what we
nowadays call crossing number of a knot, a notion
already
defined above. Tait with this notion of
crossing number, produced the first knot tables,
listing knots in order of their increasing
knottedness. If atoms had been really knots, we
would have been studying these tables instead of
the periodic table of chemical elements in our
schools.

Since the pioneering work of these physicists,
knot theory was solely investigated by
mathematicians till about ten years ago when
physicists came back to it through quantum
field theories. This brings us to modern field
theoretic interpretation of knots in three
dimensions.

\section{Abelian Chern-Simons field theory and 
knots and links}

In a field theory, the properties of a system of
infinitely many oscillators are represented collectively
by a field, $\phi(x)$ defined over all the space though
the space label $x$. An action functional is
prescribed for these fields. For example, for a
one-component scalar field $\phi (x)$, say in three
dimensional flat Euclidean space $R^3$, the action 
functional may be taken to be:

$$
S[\phi]~=~ {\frac 1 {2}} \int d^3x ~\delta^{\mu \nu}
\partial_{\mu} \phi(x) \partial_{\nu} \phi(x) ~, \nonumber 
$$

\noindent where $\mu, \nu ~=~ 1,~2,~3$ are space indices
and for $R^3$, the metric is flat $\delta^{\mu \nu}~=~
dia~( 1, 1, 1)$. For a theory defined over a general
curved three-manifold endowed with a metric $g_{\mu
\nu}$ (and its inverse $g^{\mu \nu} $), this action 
generalizes to:

$$
S[\phi]~=~ {\frac 1 {2}} \int d^3x {\sqrt {g(x)}}~g^{\mu \nu}(x) 
\partial_{\mu} \phi(x) \partial_{\nu} \phi(x) ~,  \nonumber
$$

\noindent where $ g(x) = det ~g_{\mu \nu}$.

Similarly for a vector field $A_{\mu}(x)$, the gauge
field of  Maxwell theory in three dimensions, we
write the action functional as:

$$
S[A_\mu]~=~ {\frac 1 {4}} \int d^3x {\sqrt {g(x)}}~g^{\mu
\alpha}(x) g^{\nu \beta}(x)  
 \Bigl[ \partial_{\mu} A_{\nu}(x) -  \partial_{\nu}A_{\mu}(x)
 \Bigr] 
 \Bigl[ \partial_{\alpha} A_{\beta}(x) -\partial_{\b}
A_{\a}(x)  \Bigr]  \nonumber
$$

\noindent Both these actions above are invariant under
general coordinate transformations.

Quantum field theories normally studied, like the examples
above, depend on the metric $g_{\mu \nu}$ of the three-manifold in which
the theory is defined.
The metric describes the geometric properties, such
as distances, curvature etc. But, here we
are interested in attempting a field theoretic
description of knots and links in such a way that only
their topological properties are represented. Their
size, exact shape, location etc are not of our concern.
The topological properties, unlike these, do not depend on
the metric. Thus we are seeking a field theory which is
independent of the metric. Such theories are  
called {\it topological field theories}. A simple
example of metric independent field theory is the
Chern-Simons gauge theory. Its action in the
Abelian version (with convenient
normalization) is given by:
 
\begin{equation}   
kS[A_\mu]~=~ - {\frac k {8\pi}} \int_{S^3} d^3x ~
\epsilon^{\mu \nu \alpha} A_{\mu}(x)\partial_{\nu} A_{\alpha}(x)
\end{equation}   

\noindent where $\epsilon^{\mu \nu \alpha}$ is a
completely anti-symmetric contravariant 
three-tensor density whose only nonzero
component  is $\epsilon^{1~2~3}= 1$. For
definiteness, we shall discuss this theory in
a three-manifold $S^3$. Clearly  this 
action is independent of the metric. Also it is
invariant under general coordinate transformations.
Like the Maxwell theory, this theory exhibits a gauge
invariance.

The quantum version of this theory is described by the
functional integral representing the partition
function:

\begin{equation}
 Z \ = \ \int [dA] ~e^{ikS} \label{abel}
\end{equation}

\noindent and for metric independent gauge invariant
functionals $W[A_\mu]$ of the gauge field $A_\mu (x)$,
we have the functional averages (vacuum expectation values
of the associated operators): 

\begin{equation}
\langle W \rangle \ = \ Z^{-1} \int [dA]~ W~ e^{ikS} \label{abel1}
\end{equation}

\noindent  Though the action  and gauge invariant functionals
$W$ do not depend on the metric, there are  potential
sources which can introduce  
metric dependence in these functional averages. 
The functional integration may be thought of
to be done by discretizing the space into a mesh.
Infinitely many ordinary integrals over $A_{\mu}
(x)$ at every point $x$ of the mesh are to be  done and
finally the limit of  mesh size going to zero is 
taken in some well
defined manner. This is the usual way we understand these
infinite dimensional integrals. Further, there is a gauge
invariance in the theory, which like in other gauge
theories needs to be fixed by a choice of gauge. Both
the choice of  mesh as well as  gauge fixing
condition are generically metric dependent. Thus the
gauge fixed measure of integration $[dA_\mu (x)]$ 
in a field theory defined on a curved
space, in general  depends on the metric. However,
despite these, it can be shown that various metric
dependence so conspire in this topological theory that
they cancel out without spoiling the metric
independence of the functional averages \cite{kauraj}
. 

Now let us give an explicit form of a
topological operator $W$ in this Abelian
Chern-Simons theory. Consider a link $L$ made
up of knots $K_1,~K_2,....K_s$. 
Wilson knot operator for each these knots $K_{\ell}$
is given by $exp[i~n_{\ell}\oint_{K_{\ell}}
dx^{\mu}~A_{\mu}(x)]
$ where $n_{\ell}$ is an integer measuring the
charge on the loop. Clearly these are independent of
the metric. Then the Wilson link operator is product
of all such knot operators:

\begin{equation}
W[L] \ = \ \prod_{\ell=1}^s~exp\Bigl[i~n_{\ell}\oint_{K_{\ell}}
dx^{\mu} A_{\mu}(x)\Bigr]
\end{equation}

\noindent If we expand the exponential here,
the expectation value $ \langle W[L] \rangle$ 
is given by the expectation values of the
various terms in this expansion. This is a
non-interacting theory; all these expectation
values are given in terms of the ``two-loop''
expectation values only:

\begin{equation}
\langle \oint_{K_{\ell}}dx^{\mu} A_{\mu}(x)
 \oint_{K_{m}} dy^{\nu} A_{\nu}(y)\rangle,
~~K_{\ell}  \ne   K_{m}; ~~~{\rm and}
~~~~\langle \oint_{K}dx^{\mu}A_{\mu}(x)
\oint_{K}dy^{\nu} A_{\nu}(y)\rangle \label{twoloop} 
\end{equation}  
\noindent Here in the first expression the two loops
are distinct in contrast to  the second expression 
where  both the loop integrals are along the same knot.
Clearly, these  expressions can be
easily evaluated in terms of the two-point
correlator $\langle A_{\mu}(x)~ A_{\nu}(y)
\rangle$. To do this, we can locally identify
the region containing our link with $R^3$ so
that we can use the flat metric $g_{\mu \nu}
 =  \delta_{\mu \nu}$ in this region. Then
$x^{\mu}$ and $y^{\nu}$ are the Euclidean flat
coordinates along the two knots
$K_{\ell}$ and $K_m$ respectively. This allows us
to do away with the complications connected with
the curved nature of the three-manifold $S^3$; we can
do all our calculations in flat Euclidean space
without loss of  generality. Elementary field 
theory allows us to read off
the flat space two-point correlator from the
action (subject to a gauge condition, which we
choose to be the covariant Lorentz gauge
$\delta^{\mu \nu}~\partial_\mu A_\nu \ = \ 0$
):
$$
\langle A_{\mu}(x)~ A_{\nu}(y)
\rangle \ = \ {\frac {i} {k}}~ \epsilon_{\mu \nu
\alpha}~ {\frac {(x - y)^{\alpha}} {|x - y|^3}}
$$ 
\noindent so that
\begin{eqnarray}
\langle \oint_{K_{\ell}}dx^{\mu} A_{\mu}(x)
 \oint_{K_{m}} dy^{\nu} A_{\nu}(y)\rangle
\ & =& \ {\frac {4\pi i} {k} }~{\cal L}(K_{\ell},K_{m})
~~~~~~~~~~~~~~~~~~~~~~~~~~~~~~~~~~~~~~~ \nonumber \\
{\rm where} ~~~~~~~~~~~~~~~~~~~~~~~
 {\cal L}(K_{\ell}, K_{m}) \ & =& \ {\frac 1 {4\pi}}
\oint_{K_{\ell}}dx^{\mu}\oint_{K_{m}}
dy^{\nu} \epsilon_{\mu \nu
\alpha}~ {\frac {(x - y)^{\alpha}} {|x -
y|^3}}.
\end{eqnarray} 
\noindent This double loop integral over two
distinct knots ($K_{\ell}  \ne   K_{m}$)  
is a well known topological invariant, called {\it Gauss
linking number} of the two closed curves. It
measures the number of times one knot
$K_{\ell}$ goes through the other knot $K_{m}$.
Clearly, linking number of two knots is an
integer.  For example, for
the right-handed Hopf link $H_{+}$, 

\vskip0.5cm
\centerline{\epsfbox{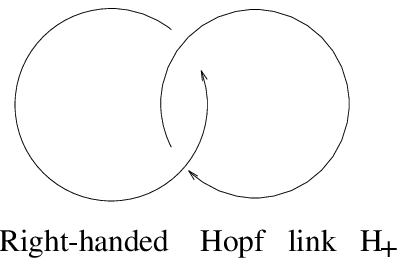}}
\vskip0.5cm
\noindent its value is $+1$.  Its value for the
mirror reflection of this link (left-handed Hopf)
is $-1$. Linking number does not
depend on the exact location of the two knots,
nor on their size or shape. It depends only on
their topological relationship with each other.
This invariant has a physical interpretation due 
to Maxwell -- in electrodynamics, it represents 
the work done to move a magnetic monopole around one
knot in three-space while an electric current
runs through the other knot.

The Abelian Chern-Simons theory also
provides a representation for yet
another simple topological quantity
associated with an individual knot
called its {\it self-linking number } and also some times 
{\it framing number} or  simply {\it framing}.
This is related to the second expectation
value given in (\ref{twoloop}) where the two loop
integrals are over the same knot. This expectation
value is to be evaluated through  a limiting
procedure: To a knot $K$ parametrized by
$x^\mu (s)$ ~$ (0 \leq s \leq L) $ along the length
of the knot by the parameter $s$, associate
another closed curve $K_f$, called its {\it frame,}
given by coordinates $y^\mu  = x^\mu (s) +
\epsilon n^\mu (s) $ where $\epsilon$ is a
small parameter and $n^\mu (s)$ is a unit
vector field normal (principal normal)
to the curve at $s$. That is, $K_f$ is the
curve $K$ displaced along the normal by a small
amount. Then the linking number of the curve
$K$ and its frame $K_f$ is called 
self-linking number ${\cal SL}(K)$ of the knot:
\begin{eqnarray}
\langle \oint_{K}dx^{\mu}A_{\mu}(x)
\oint_{K}dy^{\nu}~A_{\nu}(y)\rangle
 & = &   lim_{\epsilon \longrightarrow 0} 
\langle \oint_{K} dx^{\mu}A_{\mu}(x)
\oint_{K_f} dy^{\nu}A_{\nu}(y)\rangle  \nonumber \\
 & = & \ {\frac {4\pi i} {k}}~ {\cal L}(K,~K_f)
\ = \ {\frac {4\pi i} {k}}~{\cal SL}(K) 
\end{eqnarray}

\noindent This self-linking number is
independent of the parameter $\epsilon$
and can easily be shown to obey the following
important theorem, first proven by G. Calugareanu
almost forty years ago \cite{calu}:

{\bf Calugareanu theorem}: {\it The self-linking
number of a knot is the sum of its twist and
writhe numbers:}
\begin{eqnarray} 
{\cal SL}(K) \ & = &\  T(K) + w(K)   \\
 \nonumber \\
T(K) \ = \ {\frac 1 {2\pi}} \int_K ds 
~\epsilon_{\mu \nu \a}~ {\frac {dx^\mu}
{ds}}~n^{\nu} ~ {\frac {dn^\a} {ds}}~,   
&&w(K) \ = \ {\frac 1 {4\pi}} \int_K ds\int_K dt
~\epsilon_{\mu \nu \a}~ e^\mu~ {\frac {de^\nu}
{ds}}~ {\frac {de^\a} {dt}} \nonumber
\end{eqnarray} 

\noindent where the vector field $e^\mu $ is
given by
$$
e^\mu (s,~t) \ = \ {\frac { y^\mu(t) - x^\mu (s)}
{|y(t) - x(s)|}}   \nonumber
$$

\noindent is a map $K \otimes K \longmapsto S^2$ and
$n^\mu (s)$ is the normal vector field along the
length of the curve $K~ ( x^\mu (s), 0 \leq s \leq
L)$. The quantities $T(K)$ and $w(K)$ represent
well  defined {\it geometric} properties of the knot.
 $T(K)$ represents the {\it twist} in the knot $K$ with
reference to its frame $K_f$ and $w(K)$ is the
amount of {\it writhe} or {\it coiling} of the knot.
Clearly, the twist number and writhe number
are not necessarily integers nor
are they ambient isotopic invariants.  But
their sum, the self-linking number, is indeed
an integer and also  an ambient isotopic invariant.
This theorem  can be easily appreciated  if we recall
that stretching a coiled up telephone cord
reduces its coils but increases its twist
and loosening of a twisted cord
coils it up.  The amount of coils lost (or
gained) is exactly the same as the amount by
which the twisting is gained (or lost) so that
their sum is always unchanged. 
This theorem of Calugareanu when
applied to circular ribbon (which can be
thought of as a framed closed
curve) has been put to good use in the study of
the properties of circular polymers and
circular DNA \cite{crick}.

Notice that the  self-linking number does carry
dependence on the frame. The mathematical
concept of framing of a knot is intimately connected to
the concept of regularization in field theory. In
order to avoid the coincidence singularity in the
two-point correlator $  lim _{x \rightarrow y}
\langle A_\mu (x)~ A_\nu (y) \rangle$, we need to 
regularize it, say by point-splitting. Evaluating,
`two-loop' correlator of Eqn.(\ref{twoloop}), where the two loops
are same, we face this same divergence, which, through
framing, has been resolved by `loop-splitting'. 
Ordinarily, quantities in field theory do depend
on the regularization. Like-wise the self-linking number here
depends on the framing. But all those framing curves
enveloping around the knot, which can be continuously
deformed into each other without snapping the
knot, form a
topological class for which the self-linking number does
not change. In field theory language, framing provides
a {\it topological regularization}.

Now collecting all these pieces of information,
the expectation value of the Wilson link
operator for a link $L \ = \ (K_1, K_2,
...K_s)$ in the Abelian Chern-Simons theory on $S^3$
can be written down in terms of the linking and
self-linking (framing) numbers as:
\begin{equation}
\langle W[L] \rangle  \ = \ exp \left\{ - 
{\frac{2\pi i} {k}} 
\Bigl[ \sum_{\ell}^s n_{\ell} ^2 ~{\cal SL}(K_\ell)
+ \sum_{\ell \ne m}^s n_\ell n_m {\cal L}(K_\ell, 
K_m) \Bigr] \right\}
\end{equation}

Thus, we have indicated here how this simple field theory
does indeed, through expectation values of Wilson link
operators, provide a field theoretic interpretation of
some of the topological invariants, linking number and
self-linking number of knots and links.
Non-Abelian Chern-Simons theories are much richer in
their structure; these capture even more complex
topological properties of knots and links.

\section{Non-Abelian Chern-Simons theory as a
description of knots and links}

A non-Abelian Chern-Simons theory, instead of  being a
gauge theory of one vector field,
involves, say for gauge group $SU(2)$, three such fields,
$A_\mu ^a$ ~$(a \ = \ 1, 2, 3)$. These three are
collectively written as a matrix valued vector field
$A_\mu \ = \ A_\mu^a ~{\frac {\sigma^a} {2i}}$, where
anti-hermetian matrices $ {\frac {\sigma^a} {2i}}$ are 
the generators of the group $SU(2)$. Action functional 
defined in a
three-manifold, say $S^3$, is given by:

\begin{equation}
kS~=~ {\frac k {4\pi}} \int_{S^3} d^3x ~
\epsilon^{\mu \nu \alpha}~ tr \Bigl[ A_{\mu}(x)\partial_{\nu}
A_{\alpha}(x) + {\frac 2 3 }A_{\mu}(x) A_{\nu}(x) A_{\alpha}(x) \Bigr]
\end{equation}

\noindent Like Abelian Chern-Simons theory, this action
has no metric dependence. Besides a gauge
invariance, it is  also  invariant under general
coordinate transformations.

The topological operators are the Wilson loop (knot)
operators defined as
\begin{equation}
W_j [K]~=~ tr_j P exp \oint_K dx^\mu A_{\mu}^a T_j^a  \label{wil}
\end{equation}
\noindent for an oriented knot $K$ carrying spin $j$
representation reflected by the associated  
 representation matrices $T_j^a$ ~$(a \ = \ 1,
2, 3)$. The symbol $P$ stands for path ordering of the
exponential. This is done by breaking the length of
the knot $K$ into infinitesimal intervals of size
$dx_{m}^\mu$
around the points labeled by the coordinates
$x_{m}^\mu$ along the knot. Then path ordered
exponential is:

$$
P~ exp \oint_K dx^\mu A_{\mu}^a T_j^a \ = \ \prod_{m}
[1 + dx_{m}^\mu A_{\mu}^a (x_{m}) T_j^a]
$$

For a link $L$
made up of oriented component knots $K_1, K_2, \ldots
K_s$
carrying spin $j_1,
j_2,\ldots j_s$ representations respectively, we have
the
Wilson link operator
defined as
\begin{equation}
W_{j_1j_2\ldots j_s} [L] \ = \ \prod_{\ell=1}^s \
W_{j_\ell}
[K_\ell] \label{wil1}
\end{equation}
\noindent We are interested in the functional
averages of these operators:
\begin{equation}
V_{j_1j_2\ldots j_s}[L] \ = \ Z^{-1} \int [dA]
~W_{j_1j_2\ldots j_s} [L]
~e^{ikS}~, ~~~~~{\rm where}~~~ Z \ = \ \int [dA]
~e^{ikS} \label{jon}
\end{equation}
\noindent Here the integrands in the functional integrals are
metric
independent. So is the gauge fixed measure \cite{kauraj}.
Therefore, these expectation values depend
only on the isotopy type of the oriented link $L$ and
the set of representations $j_1, j_2\ldots j_s$ 
associated with component knots.

These expectation values can be obtained
non-perturbatively.
For example, for knots and links carrying only the spin
$1/2$ representations, Witten has shown that the link invariants
(expectation values of the associated Wilson link
operators) satisfy a simple relation. This relation
is given for three link diagrams which are identical
every where except for one crossing where they differ
in  that it is an over-crossing ($L_+$), or
no-crossing ($L_0$) or an under-crossing ($L_-$) as
shown in the figure below:
\vskip0.5cm 
\centerline{\epsfbox{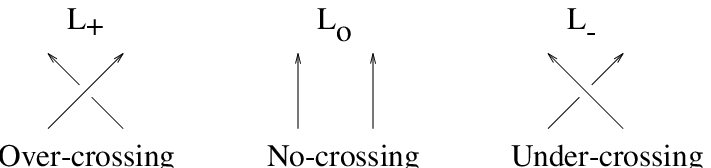}}
\vskip0.5cm 

\noindent Then the invariant for such links are
related as:
\begin{equation}
q~V_{1/2} [L_+] - q^{-1} V_{1/2} [L_-] \ = \ (q^{1/2} -
q^{-1/2}) ~V_{1/2} [L_0]
\end{equation}
\noindent where  $q$ is  a root of unity 
related  to the Chern-Simons coupling
$k$ through the relation $q = exp[2\pi i/(k + 2)]$.  
This is precisely the well known  generating skein relation for
the Jones polynomials. Indeed $V_{1/2} [L]$, which is
the expectation value of the Wilson link operator 
where every component knot carries the doublet spin
$1/2$ representation, is the one-variable
Jones polynomial. 
  
The above skein relation  is powerful enough that it
recursively yields Jones polynomial for any arbitrary link.
For example consider following three link diagrams:

\vskip0.5cm 
\centerline{\epsfbox{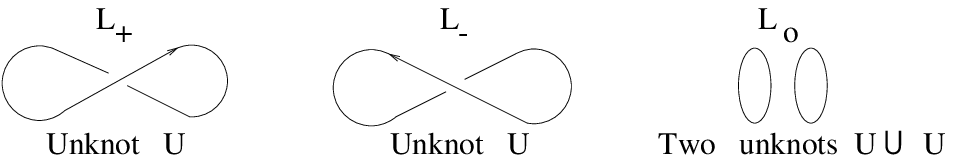}}
\vskip0.5cm 

\noindent We use an important factorization property 
of these invariant: {\it the link invariant of two distant
(disjoint) links (that is, with no mutual entanglement)
is simply the product of invariants for
the individual  links}. That is, for the link $L_0$ above,
$V_{1/2} [U \cup U]  =  (V_{1/2}[U])^2$, where
symbol $U$ represents the unknot. Then use of the
skein relation yields:
$$
q~V_{1/2} [U] - q^{-1} V_{1/2} [U] \ = \ (q^{1/2}
-
q^{-1/2})~(V_{1/2}[U] )^2
$$ 
\noindent so that spin $1/2$ invariant for an unknot
is given by: $ V_{1/2} [U]  =  q^{1/2} + q^{-1/2} $.

Next  apply the skein relation to three links, where
the $L_{+}$ is the right-handed Hopf link, $L_{-}$ is
simply the union of two (unlinked) unknots and $L_0$ is 
an unknot:

\vskip0.5cm 
\centerline{\epsfbox{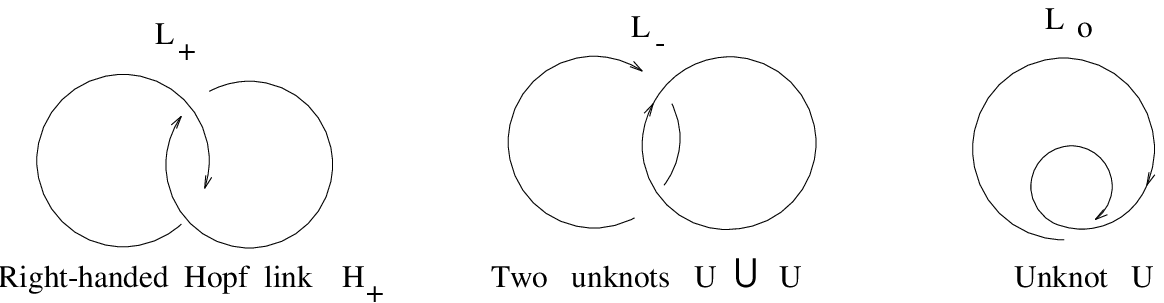}}
\vskip0.5cm 

\noindent This yields, the invariant for the right-handed
Hopf link $H_{+}$ as: $ V_{1/2}[H_{+}] \ = \ 1 + q^{-1} + 
 q^{-2} + q^{-3}$. Now use recursion relation for 
the three links:

\vskip0.5cm 
\centerline{\epsfbox{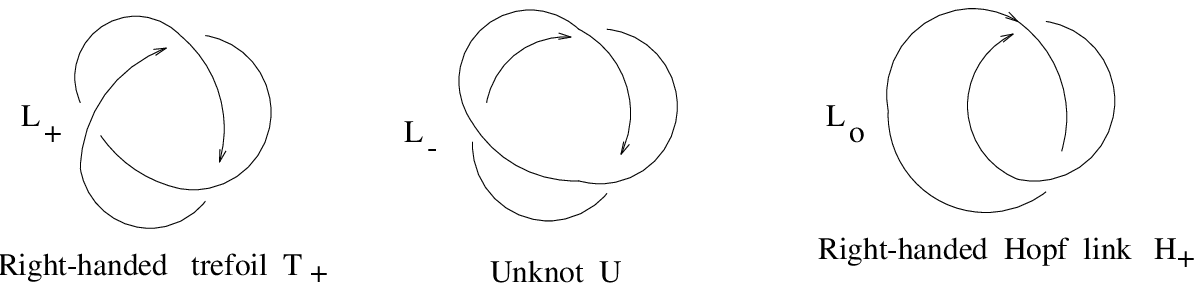}}
\vskip0.5cm 

\noindent where $L_+$ is a right-handed trefoil ($T_+$), $L_-$
is an unknot and $L_0$ is a right-handed Hopf $H_+$.
This gives us the invariant for the trefoil knot as
$ V_{1/2}[T_+] \ = \ q^{-1/2} + q^{-3/2} +
 q^{-5/2} - q^{-9/2}$. This way invariant for any
arbitrary link can be recursively obtained.

Jones polynomial is in fact the simplest of the
examples of a whole host of new link invariants that
emerge naturally from this field theory.
More general invariants  are the expectation values of
Wilson link  operators with arbitrary
spin representations placed on the knots.
The formalism  does also allow for placing different
representations on each of the component knots.
This leads to so-called {\it coloured polynomial
invariants}. Besides, instead of the gauge group
$SU(2)$,  Chern-Simons theory based on any other 
semi-simple group can be  used. These
then yield even richer spectrum of the new
invariants.

While Jones polynomial  can be
obtained by recursive use of the skein relation, other more
general invariants (for spin representations 
$j  =  1, 3/2 ...)$ can  {\it not} be
obtained in this manner.  Of course there are
generalizations  of the skein relations for an arbitrary spin 
invariants. But these  do not possess recursively complete
solutions (except for spin $1/2$ case above). 
Therefore methods had to be
developed to obtain expectation values of Wilson
operators with arbitrary representations living on
the component knots of a link. One such method in
its complete manifestations has been presented in
ref \cite{kau2}. This allows us to present  a
complete and explicit solution of the Chern-Simons
theory. This is a non-perturbative method which,
generalizing the formalism set up by Witten,
makes use of two ingredients, one from quantum field
theory and other from mathematics of braids:
\vskip0.2cm

{\it (i) Field theoretic input:}~ 
Chern-Simons theory
on a three-manifold with boundary is essentially
characterized by a corresponding two dimensional
Wess-Zumino conformal field theory on that 
boundary\cite{wit}:
\vskip0.5cm
\centerline{\epsfbox{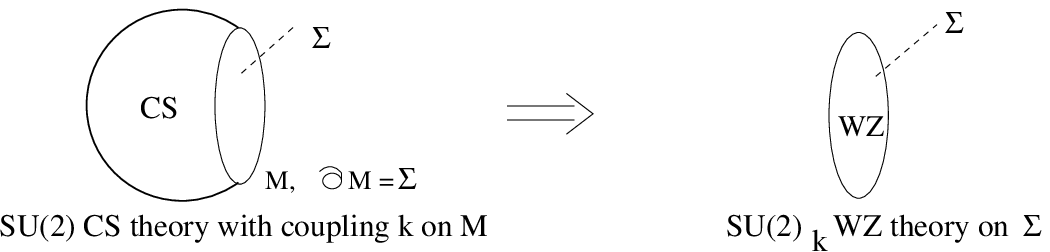}}
\vskip0.5cm
 
\noindent And Chern-Simons functional average for
Wilson
lines ending at $n$ points in the boundary is
described by
the
associated Wess-Zumino theory on the boundary
with $n$
punctures carrying the representations of the
free Wilson lines:
 
\vskip0.5cm
\centerline{\epsfbox{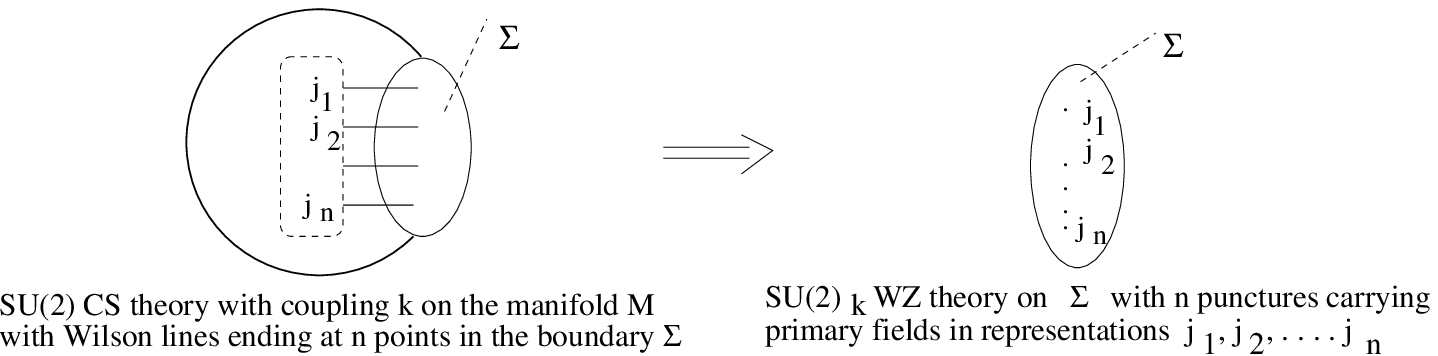}}
\vskip0.5cm
\noindent The Chern-Simons functional integral
can be
represented \cite{wit} by a vector in the Hilbert
space $\cal{H}$ associated with the space of $n$-point 
correlator of the Wess-Zumino conformal field theory 
on the boundary $\Sigma$.
In fact, these correlators provide a basis
for this boundary Hilbert space. There are more
than one possible basis. These different bases are 
related by duality of the correlators of the
conformal field theory\cite{kau2}.
 
\vskip0.2cm

{\it (ii) Mathematical input:}~ The second ingredient
used  is the close connection knots and links have with 
braids.  An $n$-braid is a collection of non-intersecting 
strands connecting $n$ points on a horizontal plane to $n$ 
points on another horizontal plane directly below
the first set of $n$ points. The strands are not
allowed to go back upwards at any point in their travel. 
The braid may be projected onto a plane with the two
horizontal planes collapsing to two parallel
rigid rods. The over-crossings and under-crossings of the 
strands are to be clearly marked.  When all the strands
are identical, we have ordinary braids.  The theory of such
braids, first developed by Artin, is well studied.
These braids form a group. However, we may wish
to orient the individual strands and further distinguish 
them by putting different colours on them. These different 
colours are represented by different $SU(2)$ spins. These 
braids, unlike braids made from unoriented identical strands, 
have a more general structure than a group. These instead
form a groupoid.  The necessary aspects of the theory of such
braids have been  presented in ref.\cite{kau2}

One way of relating the braids to knots and links is
through closure of braids. We obtain the closure of a 
braid by connecting the ends of the first, second, 
third, ..... strands from above to the ends of 
respective first, second, third, ..... strands from
below as shown in (A):

\vskip0.5cm
\centerline{\epsfbox{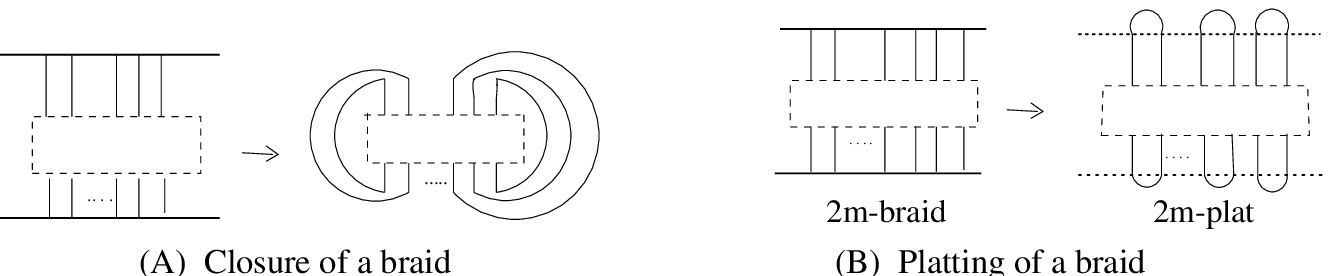}}
\vskip0.5cm

\noindent There is a theorem by Alexander\cite{alex1}
which states that {\it any knot or link can be
obtained as closure of a braid.} 
This construction of a knot or link is not unique.

There is another construction associated with
braids which relates them to knots and links.  
This is called platting.  Consider a
$2m$-braid, with pairwise adjacent strands
carrying the same colour and opposite orientations.  
Then connect the $(2i-1)$th strand with $(2i)$th from
above as well as from below.  This yields the plat of the 
given braid as shown in (B) above.  There is a theorem 
due to Birman\cite{bir} which relates
plats to links. This states that {\it a
coloured-oriented link can be represented (though
not uniquely) by the plat of an oriented-coloured 
$2m$-braid}.
 
Use of these two inputs, namely relation of
Chern-Simons theory to the boundary Wess-Zumino  conformal 
field theory and presentation of knots and links as
closures or plats of  braids leads to an
explicit, complete and non-perturbative solution
of the Chern-Simons theory. Conformal field theory  on
associated boundary gives
matrix representations  for braids and platting or closing
of a braid
corresponds to taking a specific matrix element
of these braid representations.  This then yields the expectation
value of the Wilson link operator associated with that
link. For example this invariant for an unknot $U$  carrying 
spin $j$ representation turns out to be:

$$
V_{j} [U] \ = \ [2j +1] ~~~~~~~~~~{\rm where}~~~ ~ [x] \  =  \ 
{\frac {q^{x/2} - q^{-x/2}} {q^{1/2} - q^{-1/2}}}
$$
The square bracket indicates a $q$-number.  Jones
polynomial  above corresponds to spin $j=1/2$. And for a
right-handed trefoil $T_{+}$, the invariant turns out to be:
$$
V_j [T_{+}] \ = \ \sum_{m = 0,1,2,..min(2j, k-2j)} 
~~[2m +1]~ (-)^{2j+m} ~q^{-6C_j +{\frac 3 2} C_{m}}
$$
\noindent where $C_j = j(j+1)$ is the quadratic Casimir of the
spin $j$ representation. For $j=1/2$, this expression agrees  with 
the polynomial obtained above by using the skein relation.

The link invariants calculated from the field theory
depend on the regularization used to define
the coincident loop correlators, that is, the
framing of the knots. 
The invariants  above have been obtained in a
specific framing called {\it standard framing}.
 In particular, the skein relation   for spin
$1/2$ invariants given above is in this framing.
In this framing, the self-linking (framing) number of every knot 
is zero. The invariants so obtained are unchanged 
under all the three Reidemeister moves. That is, this
yields ambient isotopic invariants. There is another 
framing choice which has been of special interest. In
this case, the frame is thought to be just vertically
above the two dimensional projection of the knot.
In this framing, known as {\it vertical framing},
Reidemeister moves II and III do leave the link invariants 
unchanged,  but Reidemeister move I changes them.

The general framework developed  provides a powerful
method of calculating knot and link invariants.
This has in the process also provided
answers to some of the open problems of knot
theory. For example, one such  problem is to  find
polynomial invariants which would discriminate
between  two chiralities of a given knot.
The invariants for the mirror reflected knots are give
by simple complex conjugation.  Up to ten crossing number, 
there are six chiral knots, $9_{42}$,  $10_{48}$,  
$10_{71}$, $10_{91}$, $10_{104}$  and $10_{125}$ (as 
listed in the knot tables in Rolfsen's book \cite{rol}) 
which are not
distinguished from their mirror images by
spin 1/2 (Jones) polynomials. Spin one 
(Kauffman/Akutsu-Wadati) polynomials
do detect the chirality of four of them, namely
$10_{48},~ 10_{91},  
~10_{104} \,~ {\rm and}\, ~ 10_{125}$.  But for
$9_{42} \, ~{\rm
and} \,~  
10_{71}$  
both Jones and Kauffman polynomials are not  changed
under chirality
transformation ${(q \rightarrow q^{-1})}$.
However,  the new spin
$3/2$ invariants are powerful enough to distinguish
these knots from their mirror images\cite{kau3} .

Another problem of knot theory that has been provided
with an answer is to do with so called  {\it mutant}
knots.  A mutant of a knot or link is obtained in the following
way: isolate a portion of the knot in such a way that it has
two strands going into and two strands leaving from
it. Scoop it out and rotate it through $\pi$ about any of 
three orthogonal axes (rotations about only two of these
are really independent). Glue it back after, if necessary, 
changing the orientations on the strands to match the free 
ends of strands of  rest of the knot to which
the free ends of the rotated portion are glued. 
This yields a mutant
of the original knot. It has been possible to prove that 
polynomial invariants obtained from
a Chern-Simons theory based on {\it any arbitrary
non-Abelian gauge group} do not distinguish 
isotopically inequivalent mutant knots\cite{kau4}.
 As an  example 
consider the following sixteen crossing mutant knots:

\vskip0.4cm
\centerline{\epsfbox{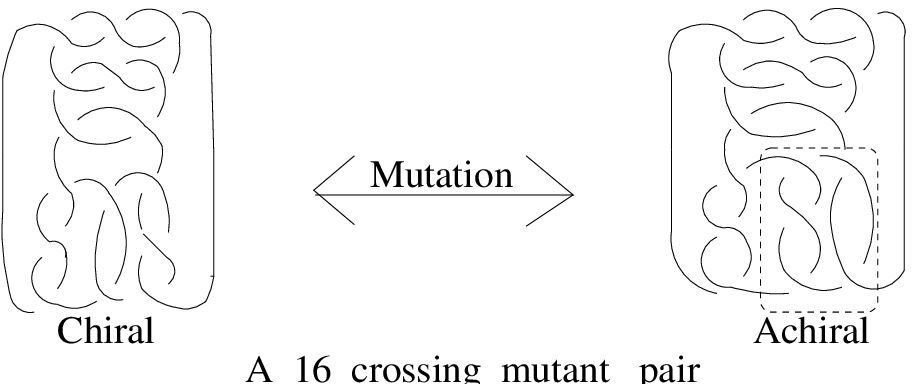}}
\vskip0.2cm
 
\noindent  The two knots are related by  a mutation of
the portion  indicated by dashed enclosure.  Like all other
mutants, the invariants
obtained from  any non-Abelian Chern-Simons theory  for them
are identical. What is of particular interest
about this pair is that one of them
is chiral, other is not. This then yields  an example of
a chiral knot whose chirality can not be detected by any of
these invariants.

The general framework developed to study knots and links
is also applicable to another 
set of gauge invariant operators  called graphs.
For $SU(2)$ Chern-Simons theory, these are the graphs
containing vertices with three legs.  The edges of the graph
between vertices carry  Wilson line operators.
More general gauge invariant operators which include
links attached to the edges of  graphs can also be
evaluated in this framework.

\section{Three-manifold invariants} 

The invariants of knots and
links in $S^3$ obtained from the Chern-Simons theory can 
be used to construct a special three-manifold 
invariant\cite{wit,res,lic, kau5}. This
provides an important tool to study  topological  properties of
three-manifolds. Starting step in this construction is a 
theorem due to Lickorish and Wallace \cite{wal,rol}:
\vskip0.2cm

{\bf Fundamental theorem of Lickorish and Wallace:}~{\it  Every
closed, orientable, connected three-manifold, $M^3$ can be
obtained by surgery on an unoriented framed knot or link $[L, ~f]$
in $S^3$.}
\vskip0.2cm

As described earlier, the framing $f$ of a link  $L$ is defined by associating
with every component knot $K_s$ of the link an
accompanying closed curve $K_{sf}$ parallel to the knot and
winding $n(s)$ times in the right-handed direction. That is,
the linking number $lk(K_s, K_{sf})$ of the component knot
and its frame (self-linking number of the
knot $K_s$) is $n(s)$.  For the construction of 
three-manifold
invariants, we use vertical  framing where 
 the frame is thought to be
just vertically above the two dimensional projection of the
knot as shown below. This is some times indicated by
putting $n(s)$ writhes in the strand making the knot or even
by just simply writing the integer $n(s)$ next to the knot
as shown below:
 
\centerline{\epsfbox{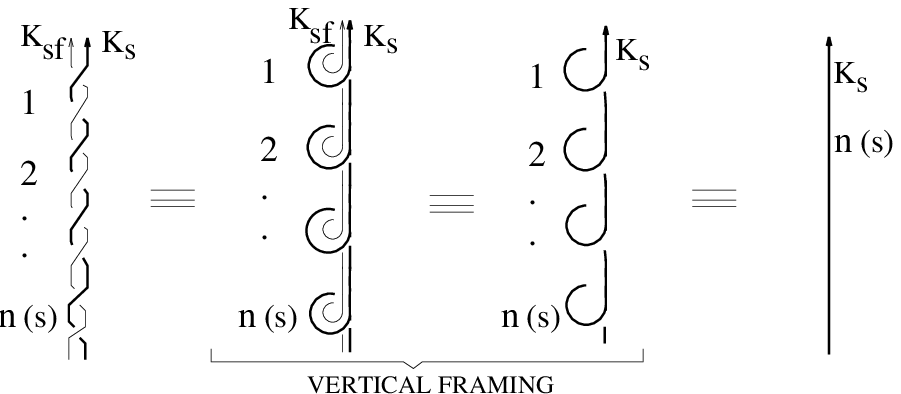}}
\vskip0.2cm
Next the surgery on a framed link $[L, f]$  made of
component knots $K_1, K_2, ....~K_r$  with framing
$f= (n(1), n(2), ....~n(r))$ in $S^3$ is performed in the following
manner. Remove a small open solid torus neighbourhood ${N_s}$
of each component knot $K_s$, disjoint from all other such
open tubular neighbourhoods associated with other component
knots. In the manifold left behind $S^3 -(N_1 \cup N_2
\cup~....~N_r)$, there are $r$ toral boundaries. On each such
boundary, consider a simple closed curve (the frame) going
$n(s)$ times along the meridian and once along the longitude
of the associated knot $K_s$. Now do a modular transformation on
such a toral boundary such that the framing curve bounds a
disc. Glue back the solid tori into the gaps. This yields a
new manifold $M^3$. The theorem of Lickorish and Wallace
assures us that every closed, orientable, connected
three-manifold can be constructed in this way.
 
This construction of three-manifolds  by surgery is not unique:
surgery on more than one framed link can yield homeomorphic
manifolds. But the rules of equivalence of framed links in
$S^3$ which yield the same three-manifold on surgery are
known. These rules are known as Kirby moves\cite{kir}.
\vskip0.2cm

{\bf Kirby calculus on framed links in $S^3$:}
Following two elementary moves (and their inverses)
generate Kirby calculus:
 
\vskip0.2cm
 
{\it Move I}. For a number of unlinked strands belonging
to the component knots $K_s$ with framing $n(s)$ going
through an unknotted circle $C$ with framing $+1$, the
unknotted circle can be removed after making a
complete clockwise twist from below in the disc enclosed
by the circle $C$:
 
\vskip0.5cm
\centerline{\epsfbox{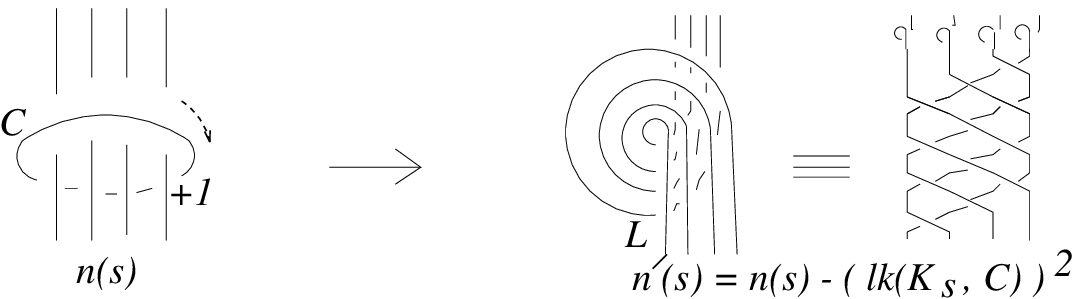}} 
\vskip0.5cm

\noindent  In the process, in addition to
introducing new crossings, the framing of the various
resultant component knots, $K'_s$ to which the affected
strands
belong, change from $n(s)$ to $n'(s)~=~n(s)- \left (
lk(K_s, C) \right )^2$.
\vskip0.2cm

{\it Move II}. Drop a disjoint unknotted circle with
framing $-1$ without any change in the rest of the
link:
 
\centerline{\epsfbox{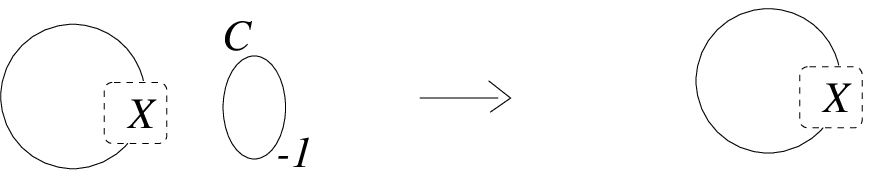}} 
\vskip0.2cm

\par  Thus Lickorish-Wallace theorem  and equivalence of
surgery under Kirby moves  reduces the theory of closed,
orientable, connected three-manifolds to the theory of
framed unoriented links via a one-to-one correspondence:
$$
\left( \matrix {Framed~ links~ in~ S^3~ modulo \cr
 equivalence~under~ Kirby ~moves} \right) ~\leftrightarrow
~
\left( \matrix
{Closed, ~orientable,~ connected~ three-\cr manifolds~
modulo~ homeomorphisms} \right)
$$

\noindent This consequently allows us to characterize
three-manifolds by the invariants of  associated
unoriented framed knots and links obtained from the
Chern-Simons theory in $S^3$. This can be done by
constructing an appropriate combination of the invariants
of the framed links which is unchanged under Kirby moves:
$$
\left( \matrix {Invariants ~of ~a ~framed ~unoriented ~link
\cr
 which ~ do~ not~ change ~under~ Kirby ~moves} \right) ~=~~
\left( \matrix
{Invariants ~ of ~associated \cr three-manifold~
} \right)
$$

One such invariant has been constructed in
ref \cite{kau5}. It is given in terms of invariants
for {\it unoriented} links obtained from $SU(2)$ 
Chern-Simons theory. The link invariants discussed
in Sec.4 above are obtained in standard framing.
These are sensitive to the relative orientations
of the component knots. Here we shall use invariants for
unoriented links in vertical framing. 
But, unlike the invariants in standard framing which exhibit 
ambient isotopic invariance, those  obtained in
vertical framing  have  only regular isotopic invariance.
That is,  in standard framing, a writhe can be stretched
(a Reidemeister move I) without affecting the link invariant, 
in vertical framing this is not so.
A link invariant in vertical framing gets changed by a phase when a writhe is
smoothed out as:
$$
\epsfbox{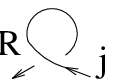} ~ =~q^{C_j}~\epsfbox{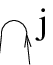},
~~~~{\rm and}~~~~~~\epsfbox{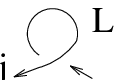}~ =~
~q^{-C_j} ~\epsfbox{Sec5fig11.eps}
$$
\noindent  where $C_j = j(j+1)$. Here we have represented the link invariant 
by the affected portion of the link. Thus, in vertical framing,
invariant for an unknot with self-linking (framing) number $+1$
or  $-1$ is related to the invariant for an unknot with zero
self-linking number as:
\begin{eqnarray*}
&&~~~~~~~~V_j  \left[~\epsfbox{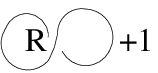} ~\right]~ =~ q^{C_j}
~V_j  \left[~ \epsfbox{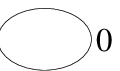} ~\right] =~q^{C_j}
[2j+1],  \\
\\       
&&{\rm and}~~~~ V_j  \left[~ \epsfbox{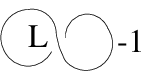} ~\right]~ =~ q^{-C_j}
V_j  \left[~ \epsfbox{Sec5fig13.eps} ~\right] ~=~q^{-C_j} [2j+1].
\end{eqnarray*}
\noindent In this framing, each right-(left-) handed
crossing in a knot introduces a self-linking number
$+1$ ($-1$). For a right-handed trefoil (self-linking number
$=3$), the invariant in this framing turns out to be:
$$
V_j[T_{+}] \ = \ \sum_{m = 0, 1, ...min(2j, k-2j)}~~ [2m +1]~
(-)^{m}~q^{-3C_{j} + {\frac 3 2}C_{m}}
$$

A three-manifold invariant is constructed from these link invariants
in vertical framing.
It has been shown that\cite{kau5}: 
{\it For a framed link $[L, f]$ with
component knots, $K_1,~K_2,~.....K_r$ and their framings
respectively as $n(1),~n(2),~.....n(r)$, the quantity}
\begin{equation}
{\hat F[L,f]}~=~ \alpha^{-\sigma[L,~f]}~
\sum_{\{j_i\}}~ \mu_{j_1} \mu_{j_2}....\mu_{j_r}~
V[L;~n(1), n(2),...n(r);~
j_1, j_2, .... j_r] \label{threeinv}
\end{equation}
\noindent {\it constructed from invariants $V$ of the
unoriented
framed link in vertical framing, is an invariant of 
the associated three-manifold
obtained by surgery on that link. Here   
the coefficients $\mu_{\ell}$ are given by 
$$
\mu_{\ell} ~=~ S_{0 \ell}~,
~~~ { where} ~~~~
S_{j \ell} ~= ~ \sqrt{2\over {k+2}}~~ sin~ {{\pi
(2j+1)~(2\ell
+1)}\over {k+2}}~ 
$$
\noindent and  $\a = exp{~3\pi ik/[4(k+2)]}$, 
and ${\sigma[L, f]}$ is the signature 
of  linking matrix $W[L, f]$: $\sigma [L, f]=
( no.
~of ~+ve ~eigenvalues~of~ W)-( no.~of~-ve~eigenvalues~
of~W) $.} The off diagonal elements of the 
linking matrix $(W[L, f])_{i j}$ are given by
linking number $lk(K_i, K_j)$ for the distinct knots
($i \neq j$) and diagonal elements ($i  = j$)
are the self-linking number (frame number)
of the knot $K_i$: $(W[L,~f])_{i i} \ = \ 
{\cal SL}(K_i) \ = \ n_i$.

It can be directly verified that  this three-manifold invariant
(\ref{threeinv}) is unchanged under Kirby moves I and II. 

\vskip0.5cm
 
{\it Explicit  examples:} ~Now computation of this invariant
for various three-manifolds is rather straight forward. We
present its value for a few three-manifolds. The
surgery description of manifolds $S^3$, $S^2 \times S^1$
and
$RP^3$ are given by an unknot with framing $+1,~0$ and $+2$
respectively.
As indicated above the invariant for an unknot with zero  framing carrying
spin $j$ representation is $ [2j + 1] = S_{0 j}/S_{0 0}$, 
where the square bracket represents the $q$-number.
Thus the invariant for $S^3$ is:

$$
{\hat F}[S^3] \ = \ {\hat F} \left[~{\epsfbox{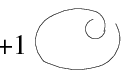}}~
\right]
\ = \  \a^{-1} ~\sum_{\ell = 0, 1/2, 1, . ..k/2} ~\mu_{\ell}
~~q^{C_{\ell}} ~~{\frac {S_{\ell 0}} {S_{0 0}}}
$$
\noindent where $\mu_{\ell} = S_{0 \ell} $ and the factor $q^{C_\ell}$
is the effect  from  framing $+1$ (one right-handed writhe). We make 
use of an identity:
$\sum_{\ell}~ S_{j \ell} ~q^{C_{\ell}}~S_{\ell m} =
\a~ q^{-C_j -C_m}~S_{j m}$~ which 
is closely related to the modular transformations of a torus.
Thus this invariant for $S^3$ is  simply:
$$
{\hat F}[S^3]~=~1~
$$

\noindent For the three-manifold $S^2 \times S^1$ (with surgery
representation as an unknot with zero framing):

$$ 
{\hat F}[S^2 \times S^1]~=~ 
{\hat F}\left[~{\epsfbox{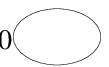}}~\right] \ = \  \sum_{\ell} \mu_{\ell}~{\frac {S_{\ell 0}}
{S_{0 0}}} \ = \ \sum_{\ell} {\frac {S_{0 \ell}S_{\ell 0}} {S_{0 0}}} \ =
\ {\frac 1  {S_{0 0}}}
$$
\noindent where orthogonality property of the $S$ matrix,
$\sum_{\ell} S_{j \ell}~ S_{\ell m} \ = \ \delta_{j m}$, 
has been used.

Next for the three-dimensional real projective space
$RP^3$ (this is an $S^3$ with antipodal points identified),
the invariant is:
$$
{\hat F}[RP^3]~=~ {\hat F}\left[~{\epsfbox{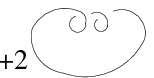}}~\right]
\ = \
\alpha^{-1} \sum_{j=0,{1\over 2},1,...{k\over 2}}{S_{0 j}
~q^{2C_j}~ S_{j 0}
\over S_{0 0}}~.
$$

A slightly more complex example we take up is the Poincare manifold
$P^3$ (also known as dodecahedral space or Dehn's homology sphere). 
It is a   homology three-sphere  given by the set
of points $(u, v, w)$  in complex $3$-space such that $u^2 + v^3 + w^5 = 0$
and $|u|^2 + |v|^2 + |w|^2 = 1 $. Its surgery presentation  is given \cite{rol} by
a right-handed trefoil knot with framing $+1$:
\vskip0.2cm
\centerline{\epsfbox{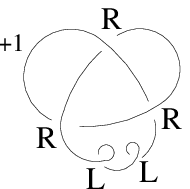}}
\vskip0.2cm
\noindent Notice, each of threeright-handed crossings 
introduces $+1$ linking number between the trefoil knot  and its
vertical
framing, and each of  two left-handed writhes
contributes $-1$ so
that the total frame number of this knot is $+1$.
Now using the knot invariant
for trefoil in vertical framing given above,  
the invariant for this three-manifold can easily be written down:
$$
{\hat F}[P^3]~=~{\alpha}^{-1}~~\sum_{j = 0,,{1 \over
2}, 1,~...{k \over 2}}~S_{0 j}q^{-2C_j}~\sum_{m = 0, 1, ...min(2j, k-2j)}
 ~(-)^{m}~[2m +1] q^{-3C_{j} +{\frac {3C_m} 2}}~
$$
\noindent The two left-handed writhes introduce a factor of
$q^{-2C_j}$.

The invariant $\hat F $ for a manifold $M^3$
constructed above is same, up to
a normalization, as the partition function of
an $SU(2)$ Chern-Simons theory on that manifold\cite{kau5,
naik}:

\begin{equation}
Z[M^3] = {\hat F}[M^3] ~ S_{00}.
\end{equation}

\noindent Generally, it is rather difficult to obtain the
Chern-Simons partition function for a given
three-manifold $M^3$ directly. But, the formulae 
above, make its computation through $\hat F$ rather 
easy.
 
The three-manifold invariant presented here is given in
terms of  link invariants from
 $SU(2)$ Chern-Simons theory. It is clear that a similar
construction can be done with link invariants from
Chern-Simons gauge theories based on
other semi-simple groups. This would yield a new
method of obtaining the partition function of such
Chern-Simons theories.
 
Next question we may ask is: ~Is this three-manifold
invariant complete? ~Two  manifolds $M$
and $M'$ for which the invariants ${\hat F}[M]$ and ${\hat
F}[M']$
are different can not be homeomorphic to each other.
But the converse is not always true;  for two arbitrary
manifolds, the invariants  need not be always different.
Recall the invariants obtained from
Chern-Simons theory  for mutant knots are not
distinct. Hence, manifold obtained by surgery on 
topologically inequivalent mutant
knots can not be distinguished by this three-manifold
invariant.

\section{Perturbative non-Abelian Chern-Simons theory}

Though Chern-Simons theories have been solved exactly and
non-perturbatively as discussed above, perturbative analysis
of these theories are also rich in their mathematical 
structure. If we expand the expectation value of the 
Wilson loop operator associated with a knot as  a 
perturbative power series in the coupling constant, 
the coefficients of such an expansion have a deep 
mathematical meaning. These on their own
are topological invariants characteristic of  the knot. 

Last decade  has  also witnessed enormous research 
activity in direct perturbative calculations in Chern-Simons
gauge field theory \cite{lab}. By simple power counting 
this theory is superrenormalizable. There are divergences, 
which need to be regularized. The effective coupling 
constant $k$ does in general depend on the regularization. 
In a class of regularizations, a shift
in the coupling constant takes place: ~$k \rightarrow k + 2$ for
$SU(2)$ theory. This
shift is consistent with the effective coupling in the
non-perturbative studies of the theory.

It is very easy to see that the first order contribution
to the vacuum expectation value of the  Wilson loop operator 
for a knot is the self-linking number of  knot up 
to some group theoretic factors. This is so because at 
this order, the theory reduces essentially to
Abelian Chern-Simons theories.  Topological regularization of the
coincident loop integrals through framing as discussed in
Sec.3 earlier, leads to this result. Higher
order contributions to the expectation value
of a Wilson loop operator  in an $SU(2)$
Chern-Simons theory yield the 
famous {\it Vassiliev invariants}. These were first
introduced by V.A. Vassiliev in 1990 from a
totally different mathematical framework
involving a study of the space of all smooth maps
of $S^1$ into $S^3$. These maps have different
types of singularities. According to
the type of singularities, this space of  maps 
divides into classes, each of which corresponds 
to a knot type. These classes are characterized
by the families of invariants characterizing 
the knot\cite{vass}.

Perturbative studies of Chern-Simons theory have provided
new insights into the theory of Vassiliev  invariants.
In a gauge theory, perturbative calculations are
to be performed in a definite gauge. 
Calculations in the Landau gauge \cite{gua} lead to
covariant integral representations of
Vassiliev  invariants, also known as
configuration space integrals first developed
by Bott and Taubes in 1994 \cite{bott}. Another integral 
representation of the Vassiliev  invariants 
was introduced by M. Kontsevich in 1993 \cite{kon}. This
corresponds to perturbative  calculation of the
Chern-Simons theory in light-cone gauge \cite{catt}.   
It  is rather very difficult to realize
that these
two integrals  represent the same invariant.
However, from a field theoretic point of view,
this is simply a consequence of gauge invariance.
Calculations in the temporal gauge have yielded
yet another formulation of these
invariants, leading  to combinatorial
formulae for them \cite{lab2}.  

\section{Gravity and Chern-Simons theory}

While Chern-Simons theories have provided a powerful
framework for theory of knots, these field theories are
also of direct relevance in physics. For example there
is an intimate relationship between these field theories 
and  three dimensional gravity which is also a topological 
field theory. In fact two copies of $SU(2)$ Chern-Simons 
theories represent gravity in Euclidean three-space with a 
negative cosmological constant\cite{wit89}.  To see this, 
just consider the partition function of  two $SU(2)$ 
Chern-Simons theories recast in terms of an
$SL(2, C)$ Chern-Simons theory as:

\begin{eqnarray*}
Z \ = \ \int [dA, d{\bar A}]~ exp~ \Bigl\{ ~{\frac {ik} {8\pi}}
\int_{M^3} d^3x~ \epsilon^{\mu \nu \alpha}&\Bigl[~tr~(A_{\mu}\partial_{\nu}
A_{\alpha} + {\frac 2 3 }A_{\mu} A_{\nu} A_{\alpha})~~&\Bigr. \Bigr.  \\
&~~~~~~-~\Bigl. \Bigl. tr~( {\bar A}_{\mu}\partial_{\nu} {\bar A}_{\alpha} +
{\frac 2 3 }{\bar A}_{\mu} {\bar A}_{\nu}
{\bar A}_{\alpha})&\Bigr] ~ \Bigr\} 
\end{eqnarray*}

\noindent where $A$ is an the $SL(2, C)$
gauge field and $\bar A$ its conjugate. This partition function
is square of two $SU(2)$ partition functions: $Z_{SL(2,C)} =
|Z_{SU(2)}|^2$.  Make a  change of variables $ A  =  \w + 
ie/\ell$ and ${\bar A} = \w -ie / \ell$,  where $\w$ and 
$e$ are the gravitational spin connection and triad respectively.  
Writing  $kS[A] = {\frac {k} {8 \pi}} \int d^3x  
\epsilon^{\mu \nu \alpha}~ tr[A_{\mu}\partial_{\nu}
A_{\alpha}+ {\frac 2 3 }A_{\mu} A_{\nu} A_{\alpha}]$, this then relates the action of these two Chern-Simons theories
to Einstein-Hilbert action for three dimensional gravity:

\begin{equation}
ik \Bigl(S[A] -S[{\bar A}]\Bigr) \ = \ {\frac 1 {16\pi G}} 
\int_{M^3} d^3x {\sqrt {g}} \Bigl(R
+ {\frac 2 {\ell^2}}\Bigr)
\end{equation}  
\noindent where the cosmological constant $= - 1/ {\ell^2} $ is negative 
and  the Chern-Simons coupling is related to the gravitational 
coupling as $k =\ell/(4G)$.

This is closely related to another development in
gravity. Three-dimensional gravity has
a lattice formulation, first introduced by
 G. Ponzano and T. Regge in 1968 \cite{pon}.
 Here the three-manifold is
decomposed into simplices. Each three-simplex is a
tetrahedron. To each edge of the tetrahedron, a
half-integral spin $j$, called its {\it colour},
 is assigned so 
that its length is given by $\sqrt  {j(j+1)}$. 
The spins on the three edges of each
triangular face satisfy the triangular angular momentum
inequality relations. 
The gravitational partition function
is constructed in terms of  Racah-Wigner six-$j$
symbols for each tetrahedron in the simplicial
decomposition of the manifold. For large spins,
the six-$j$ symbols reproduce the ordinary
gravitational action. Ponzano-Regge partition
function suffers from a problem: it diverges 
as all possible spin values are allowed to live on
the edges. This, therefore requires a regularization.
A slightly more
complex generalization of this lattice gravity
model, which also provides a regularization,
is related to a model first introduced by 
V.G. Turaev and O.Y. Viro \cite{turvir}. It 
replaces the ordinary $6$-$j$
symbols by their $q$-deformed analogues (with $q$
as a root of unity). 
For large spin values,
 the  $q$-six-$j$ symbol  can be shown to 
give Regge action for a
tetrahedron and represents Euclidean gravity action 
with a negative cosmological constant. 
The Turaev-Viro model would then be a quantum
description of this three dimensional gravity.

For a triangulation of the
three-manifold in terms of tetrahedra labeled by
$t$ and colouring $j_e$ of its edges labeled by
$e$, Turaev-Viro partition function for a manifold without 
boundary is given by the formula:
\begin{eqnarray}
Z_{TV} ~ = &&\sum_{colourings~ j_e \leq k/2}
~~\prod_{vertices} ~{\frac 1 {\Lambda}}
~~\prod_{edges~ e} ~(-1)^{2j_e}~ [2j_e +1] \nonumber\\
&& \times ~\prod_{tetrahedra~ t}~ exp\Bigl(-i\pi \sum_{i} j_i
(t) \Bigr) ~\left\{ \begin{array}{ccc}

j_1(t) & j_2(t) & j_3(t) \\

j_4(t) & j_5(t) & j_6(t) \\
\end{array}
\right \}_q 
\end{eqnarray} 
\noindent The  the square brackets indicate a $q$-numbers, 
and curly brackets 
represent a $q$-$6j$ symbol. The deformation parameter $q$ is 
related to the Chern-Simons coupling by $q = exp[2\pi i/(k
+2)]$ and $\Lambda = 
-2(k +2) / (q^{1/2} -q^{-1/2})^2 = {(S_{00})}^{-2}$. 
This partition function is naturally regularized and
finite due the restriction on the spins living on the
edges ($j_e \leq k/2$) introduced by the fact that the
deformation parameter is a root of unity.
Further this  partition function 
can be shown to be exactly square of an $SU(2)$ Chern-Simons 
partition function, $Z_{TV} = |Z_{SU(2)}|^2$. This provides
yet another representation for the Chern-Simons partition
function.

Notice that the integration measure in the  partition
function of two Chern-Simons theories above
is $[dA,d\bar{A}]$, whereas for the gravity
partition
function, it is $[de, d\omega ]$. Since $ A = \omega +
ie/ \ell$ and $ {\bar A} = \omega - ie / \ell$,
the relation between the two
involves $1/ \ell$ factors as the Jacobian. In fact in more
exact treatment, it becomes clear that  the Jacobian
for this change of variables introduces exactly 
a factor of $\Lambda$ for every vertex of the triangulation,
so that the gravity partition
function is just the Turaev-Viro partition function without
the $1/ \Lambda$ factors:
\begin{eqnarray}
Z_{grav}~=&&\sum_{colourings~j_e \leq k/2}
~~~\prod_{ edges~ e} ~(-1)^{2 j_e} ~[2j_e + 1] \nonumber \\
&& \times \prod_{tetrahedra ~ t} ~\exp{\Bigl(-i\pi \sum_{i} 
j_{i}(t)\Bigr) }
~~\left\{ \begin{array}{ccc}
j_1(t) & j_2(t) & j_3(t) \\
j_4(t) & j_5(t) & j_6(t) \\
\end{array}
\right \}_q \label{grpf}
\end{eqnarray}

\noindent For a manifold with boundary, 
this expression 
has  additional factors of ~$\exp{(i \pi
j_b)}~ \sqrt{[2 j_b + 1]}$~ for every  boundary edge
with a spin $j_b$. This partition function then
is a functional of the boundary
triangulation and spins of  edges on the boundary. 

There are many interesting questions which can be addressed
in this framework for three-dimensional gravity. Some
of these are: how does a black hole look in this 
formulation? What is its entropy? Analysis shows that 
a black hole (Banados-Teitelboim-Zanelli black hole) 
is given by a solid torus. Its horizon is given by the 
longitudinal circle at the core of this solid torus. 
The possible states associated with this black hole are 
the states associated with different triangulations of 
the black hole manifold, with the restriction that the 
longitudes have same circumference.  It can be shown that 
correct semi-classical behaviour of entropy is reproduced 
by states corresponding to all possible triangulations of  
such an  Euclidean black hole \cite{sun}. The dominant 
contribution comes from the states at the horizon.

Chern-Simons theories  have also played an important  
role in non-perturbative formulation of canonical 
quantum gravity 
in four dimensions \cite{ash}. In this approach, 
the physical states are given by spin-networks with
associated graphs in three-space, where edges are 
labeled by $SU(2)$ spins (colours) and
vertices are given by interwinning operators.
Quantum mechanical operators corresponding to 
lengths, areas and volumes all have discrete spectrum.
It can be argued that the boundary
degrees of freedom of a black hole, say Schwarzschild 
black hole, in this four dimensional theory can be described 
by a Chern-Simons theory\cite{smol,ash1}. The action embodying
the appropriate boundary conditions on the black hole 
horizon consists of, in addition to the Einstein-Hilbert 
action (in suitable variables), an $SU(2)$ Chern-Simons 
gauge theory living on  a coordinate chart of a constant 
finite cross-sectional area on the horizon. The 
Chern-Simons coupling $k$ is proportional to  this 
constant cross-sectional area.  As the fundamental
quantum excitations are polymer like, the horizon
area is generated by the punctures where these
spin-polymers pierce it.  A bulk polymer state
that gives the horizon its area in this manner
has to be compatible with the surface states on
the horizon itself. These boundary states are
described by a quantum $SU(2)$ Chern-Simons
theory on the horizon.  That is,  the space of 
these boundary degrees of freedom is given by the space of 
states of Chern-Simons theory on a three-manifold 
with an $S^2$ boundary with finitely many punctures 
on which spins live. 
The entropy of the black 
hole emerges from these boundary states. For large areas, 
where essentially $U(1)$ subgroup of $SU(2)$ contributes, 
the entropy is calculated by counting these
states. Their number  grows  exponentially with
horizon area yielding the semi-classical Bekenstein-Hawking  
expression for black hole entropy\cite{ash1}. For finite 
areas, full $SU(2)$ counting has to be done.  This has 
been done  by exploiting the relation
between the boundary states of  Chern-Simons theory 
and the space of conformal blocks of  associated 
Wess-Zumino conformal field theory on the boundary 
two-sphere, a relationship which played a crucial role
in obtaining the link invariants in Sec.4. This yields an
{\it exact} formula for entropy of a non-rotating
black hole which  for large
areas reproduces  the semi-classical formula, but for
finite areas goes beyond the Bekenstein-Hawking 
result\cite{kau6}.

\section{Summary and Concluding remarks}

We have made an attempt here to indicate how  quantum field 
theories, which have been successfully used to 
describe physics of fundamental interactions of
Nature, can also be used to study 
geometry and topology of low dimensional manifolds. These
developments not only provide  new insights into old 
problems of topology of these manifolds but also 
have been responsible 
for profoundly interesting new mathematical results. 
These developments make use of many of the recent 
developments in quantum field theories.  The interaction 
between quantum physics and mathematics has enriched 
both.

Chern-Simons
gauge field theory, a topological quantum field theory,
provides a powerful framework for  modern theory of 
knots and links in any three-manifold. This is one of 
the rare quantum field theories which can be explicitly and 
non-perturbatively solved. While Abelian Chern-Simons
theory provides a simple description of linking
and self-linking numbers of a link, non-Abelian
theories are even richer. For every representation
of any non-Abelian gauge group, there is  a new link 
invariant.  Jones polynomial  associated with spin 
$1/2$ representation in an $SU(2)$ Chern-Simons theory, 
is the simplest example of 
such link invariants. Even more general invariants 
({\it coloured invariants}) are obtained if we place 
different representations on the component knots. 
The framework is rich enough to discuss the knots 
and links not only in simple manifold like 
$R^3$ or $S^3$, but any arbitrary three-manifold. 
Chern-Simons partition function is a particularly 
interesting three-manifold invariant for which
a simple and efficient computational method is 
available now. 
Perturbative studies of Chern-Simons theory have 
given a new framework for describing
Vassiliev invariants. 

In the process of developing this framework for 
knot theory, new representations
of braids also have been obtained. The close 
connection that braids have with Yang-Baxter equation, 
has provided methods of obtaining a variety of new 
exactly solvable  two-dimensional statistical 
mechanical models in physics\cite{kau5}. These models 
are the higher vertex generalizations of the six-vertex 
model of  Lieb and Wu  and 19-vertex model of 
Zamolodchikov and Fateev.

Chern-Simons field theories are also of direct interest 
in other areas of physics. One area where these have 
found profound application is quantum
gravity. Three-dimensional gravity with a negative 
cosmological constant, itself a topological field theory, 
is essentially described by two
$SU(2)$ Chern-Simons theories. Micro-states  
of a black hole in the four dimensional spin-polymer
gravity can also be modeled
by a  Chern-Simons theory. This allows an exact 
computation of black hole
entropy going beyond the semi-classical result. 
These calculations so far  have been done for
non-rotating black holes only. These need to
be extended for  charged and rotating black holes, 
which requires certain amount of technical work. 
Further, while an exact  formula for quantum entropy
of a non-rotating black hole has been derived, a similar 
exact formula for the expectation value of the
area operator in the Chern-Simons approach
is not known. 
Also, a satisfactory understanding of Hawking radiation in
this picture is yet to be developed.

String theory is another interesting framework in which
black hole entropy has been analyzed in recent
times. Though it provides a fundamental quantum
description, unfortunately, calculations in this theory
can be done for extremal or near extremal black
holes only. These despite their mathematical interest
are not astrophysically realistic. In particular,
black holes of interest such as a Schwarzschild black hole
are not generally amenable to analysis in this
approach. Also supersymmetry plays an important role 
in the string picture. In contrast, modeling of
micro-states of a black hole by an effective
Chern-Simons theory is not limited by the
constraint of extremality or near extremality.
This framework handles the curved
 geometry of the black hole directly without invoking
supersymmetry.

There are other topological quantum field theories also. One particularly
interesting class is  so called {\it cohomological field theories}.
These are the field theoretical interpretations of 
four-manifold invariants obtained by S. Donaldson in 1983. 
His work is an example of developments in mathematics which have made
critical use of some of the notions of
 physics \cite{donal}. His theory provides
an understanding of the geometry in four dimensions
through self-dual and
anti-self-dual Yang-Mills gauge fields known to
physicists as `instantons and anti-instantons'.
Five years later, E. Witten  provided a quantum field
theoretical framework for Donaldsons's work in terms of a 
four dimensional topological  Yang-Mills gauge field theory\cite{wit1}. 
 This field theory  has certain  kind
of twisted supersymmetry. Donaldson invariants are given as the 
correlation functions in this field theory. In recent years, 
this area has registered
even further boost through the work of Seiberg and
Witten \cite{sei}. These developments use the powerful
electric-magnetic duality to relate the cohomological 
field theory based on gauge group $SU(2)$ to that based
on $U(1)$. This brings in completely new insights  into
this area and makes calculation of Donaldson four manifold
invariants rather easy.

\end{document}